\documentclass[10pt,a4paper]{article}

\usepackage[]{hyperref}
\usepackage{amsmath, amsthm, amssymb}

\title{Does gravity induce wavefunction collapse? \\An examination of Penrose's conjecture}

\author{Shan Gao\thanks{Institute for the History of Natural Sciences, Chinese Academy of Sciences, Beijing 100190, P. R. China. E-mail:  \href{mailto:gaoshan@ihns.ac.cn}{gaoshan@ihns.ac.cn}.}}

\begin{document}
\maketitle

\begin{abstract}\noindent 

According to Penrose, the fundamental conflict between the superposition principle of quantum mechanics and the principle of general covariance of general relativity entails the existence of wavefunction collapse, e.g. a quantum superposition of two different space-time geometries will collapse to one of them due to the ill-definedness of the time-translation operator for the superposition. In this paper, we argue that Penrose's conjecture on gravity's role in wavefunction collapse is debatable. First of all, it is still a controversial issue what the exact nature of the conflict is and how to resolve it. Secondly, Penrose's argument by analogy is too weak to establish a necessary connection between wavefunction collapse and the conflict as understood by him. Thirdly, the conflict does not necessarily lead to wavefunction collapse. For the conflict or the problem of ill-definedness for a superposition of different space-time geometries also needs to be solved before the collapse of the superposition finishes, and once the conflict has been resolved, the wavefunction collapse will lose its physical basis relating to the conflict. In addition, we argue that Penrose's suggestions for the collapse time formula and the preferred basis are also problematic.

\end{abstract}


\vspace{6mm}

In standard quantum mechanics, it is postulated that when the wave function of a quantum system is measured by a macroscopic device, it no longer follows the linear Schr\"{o}dinger equation, but instantaneously collapses to one of the wave functions that correspond to definite measurement results. However, this collapse postulate is not satisfactory, as it does not explain why and how the wave function collapses during a measurement. There have been various conjectures on the origin of wavefunction collapse, and the most promising one is perhaps Penrose's gravity-induced collapse argument (Penrose 1996). In this paper, we will present a critical analysis of Penrose's intriguing conjecture.

It seems very natural to guess the collapse of the wave function is induced by gravity. The reasons include: (1) gravity is the only universal force being present in all physical interactions; (2) gravitational effects grow with the size of the objects concerned, and it is in the context of macroscopic objects that linear superpositions may be violated. The gravity-induced collapse conjecture can be traced back to Feynman (1995). In his \emph{Lectures on Gravitation}, Feynman considered the philosophical problems in quantizing macroscopic objects and contemplates on a possible breakdown of quantum theory. He said, ``I would like to suggest that it is possible that quantum mechanics fails at large distances and for large objects, …it is not inconsistent with what we do know. If this failure of quantum mechanics is connected with gravity, we might speculatively expect this to happen for masses such that $GM^2/\hbar c=1$, of $M$ near $10^{-5}$ grams."\footnote{It is worth noting that Feynman considered this conjecture even earlier at the 1957 Chapel Hill conference (DeWitt and Rickles 2011, ch.22).}

Feynman's suggestion was later investigated by several authors (e.g. K\'{a}ro lyh\'{a}zy 1966; K\'{a}rolyh\'{a}zy, Frenkel and Luk\'{a}cs 1986; Di\'{o}si 1984, 1987, 1989; Penrose 1981, 1986, 1989, 1994, 1996, 1998, 2000, 2002, 2004). In particular, Penrose (1996) proposed a detailed gravity-induced collapse argument, and the proposal is a `minimalist' one in the sense that it does not aspire to a more complete dynamics. The argument is based on a fundamental conflict between the superposition principle of quantum mechanics and the principle of general covariance of general relativity. The conflict can be seen by considering the superposition state of a static mass distribution in two different locations, say position A and position B. On the one hand, according to quantum mechanics, the valid definition of such a superposition requires the existence of a definite space-time background, in which position A and position B can be distinguished. On the other hand, according to general relativity, the space-time geometry, including the distinguishability of position A and position B, cannot be predetermined, and must be dynamically determined by the position superposition state. Since the different position states in the superposition determine different space-time geometries, the space-time geometry determined by the whole superposition state is indefinite, and as a result, the superposition state and its evolution cannot be consistently defined. In particular, the definition of the time-translation operator for the superposed space-time geometries involves an inherent ill-definedness, leading to an essential uncertainty in the energy of the superposed state. Then by analogy Penrose argued that this superposition, like an unstable particle in quantum mechanics, is also unstable, and it will decay or collapse into one of the two states in the superposition after a finite lifetime. 

Moreover, Penrose (1996) suggested that the essential energy uncertainty in the Newtonian limit is proportional to the gravitational self-energy $E_{\Delta}$ of the difference between the two mass distributions\footnote{Penrose's Newtonian expression for the energy uncertainty has been generalized to an arbitrary quantum superposition of relativistic, but weak, gravitational fields (Anandan 1998).}, and the collapse time, analogous to the half-life of an unstable particle, is 

\begin{equation}
T \approx \hbar/E_{\Delta}.
\label{PCT}
\end{equation}

\noindent This criterion is very close to that put forward by Di\'{o}si (1989) earlier\footnote{In Di\'{o}si's (1989) collapse model, the increase of energy induced by wavefunction collapse is too large to be consistent with experiments. This problem was pointed out and solved by Ghirardi, Grassi and Rimini (1990).}, and it is usually called the Di\'{o}si-Penrose criterion. Later, Penrose (1998) further suggested that the preferred bases (i.e. the states toward which the collapse tends) are the stationary solutions of the so-called Schr\"{o}dinger-Newton equation within Newtonian approximation.

Now let's examine Penrose's gravity-induced collapse argument in detail. The crux of the argument is whether the conflict between quantum mechanics and general relativity requires that a quantum superposition of two space-time geometries must collapse after a finite time. We will argue in the following that the answer seems negative. First of all, although it is widely acknowledged that there exists a fundamental conflict between the superposition principle of quantum mechanics and the principle of general covariance of general relativity, it is still a controversial issue what the exact nature of the conflict is and how to resolve it. The problem is often referred to as the `problem of time' in various approaches to quantum gravity (Kucha\v r 1992; Isham 1993; Isham and Butterfield 1999; Kiefer 2007; Anderson 2012). It seems not impossible that the conflict may be solved by reformulating quantum mechanics in a way that does not rely on a definite spacetime background (see, e.g. Rovelli 2004, 2011).

Secondly, Penrose's argument by analogy seems too weak to establish a necessary connection between wavefunction collapse and the conflict between general relativity and quantum mechanics. Even though there is an essential uncertainty in the energy of the superposition of different space-time geometries, this kind of energy uncertainty is different in nature from the energy uncertainty of unstable particles or unstable states in quantum mechanics (Gao 2010). The former results from the ill-definedness of the time-translation operator for the superposed space-time geometries, while the latter exists in a definite spacetime background, and there is a well-defined time-translation operator for the unstable states. Moreover, the decay of an unstable state (e.g. an excited state of an atom) is a natural result of the linear quantum evolution, and the process is not random but deterministic. In particular, the decay process is not spontaneous but caused by the background field constantly interacting with the unstable state, e.g. the state may not decay at all when being in a very special background field with bandgap (Yablonovitch 1987). By contrast, the hypothetical decay or collapse of the superposed space-time geometries is spontaneous, nonlinear and random. In short, there exists no convincing analogy between a superposition of different space-time geometries and an unstable state in quantum mechanics. Accordingly, one cannot argue for the collapse of the superposition of different space-time geometries by this analogy. Although an unstable state in quantum mechanics may decay after a very short time, this does not \emph{imply} that a superposition of different space-time geometries should also decay - and, again, sometimes an unstable state does not decay at all under special circumstances. To sum up, Penrose's argument by analogy only has a very limited force, and it is not strong enough to establish a necessary connection between wavefunction collapse and the conflict between quantum mechanics and general relativity\footnote{In our opinion, Penrose also realized the limitation of the analogy and only considered it as a plausibility argument.}.

Thirdly, it can be further argued that the conflict between quantum mechanics and general relativity does not necessarily lead to wavefunction collapse. The key is to realize that the conflict also needs to be resolved before the wavefunction collapse finishes, and when the conflict has been resolved, the wavefunction collapse will lose its physical basis relating to the conflict. As argued by Penrose (1996), a quantum superposition of different space-time geometries and its evolution are both ill-defined due to the fundamental conflict between the principle of general covariance of general relativity and the superposition principle of quantum mechanics, and the ill-definedness requires that the superposition must collapse into one of the definite space-time geometries, which has no problem of ill-definedness. However, the wavefunction collapse seems too late to save the superposition from the ``suffering" of the ill-definedness during the collapse. In the final analysis, the conflict or the problem of ill-definedness needs to be solved \emph{before} defining a quantum superposition of different space-time geometries and its evolution. In particular, the hypothetical collapse evolution of the superposition also needs to be consistently defined, which again indicates that the wavefunction collapse does not solve the problem of ill-definedness. On the other hand, once the problem of ill-definedness is solved and a consistent description obtained, the wavefunction collapse will lose its connection with the problem\footnote{Note that if the problem of ill-definedness cannot be solved in principle for the superpositions of very different space-time geometries, then wavefunction collapse may be relevant here. Concretely speaking, if the superpositions of very different space-time geometries cannot be consistently defined in principle, then these superpositions cannot exist and must have collapsed into one of the definite space-time geometries before being formed from the superpositions of minutely different space-time geometries. In this case, the large difference of the space-time geometries in the superposition will set an upper limit for wavefunction collapse. Though the limit may be loose, it does imply the existence of wavefunction collapse. However, this possibility seems very small.}. Therefore, contrary to Penrose's expectation, it seems that the conflict between quantum mechanics and general relativity does not entail the existence of wavefunction collapse.

Even though Penrose's gravity-induced collapse argument may be problematic, it is still possible that wavefunction collapse is a real physical process\footnote{It has been recently argued that the de Broglie-Bohm theory and the many-worlds interpretation seem inconsistent with the meaning of the wave function derived based on protective measurements (Gao 2011). If the argument is valid, then the result strongly suggests that wavefunction collapse is a real physical process.}. Moreover, Penrose's Eq. (\ref{PCT}) can also be assumed as it is, and numerical estimates based on the equation for life-times of superpositions indeed turn out to be realistic (Penrose 1994, 1996). Therefore, Penrose's suggestions for the collapse time formula and the preferred basis also deserve to be examined as some aspects of a phenomenological model.

To begin with, let's analyze Penrose's collapse time formula, Eq. (\ref{PCT}), according to which the collapse time of a superposition of two mass distributions is inversely proportional to the gravitational self-energy of the difference between the two mass distributions. As we have argued above, there does not exist a precise analogy between such a superposition and an unstable state in quantum mechanics, and gravity does not necessarily induce wavefunction collapse either. Thus this collapse time formula, which is originally based on a similar application of Heisenberg's uncertainty principle to unstable states, will lose its original physical basis. In particular, the appearance of the gravitational self-energy term in the formula is in want of a reasonable explanation (see below). In fact, it has already been shown that this gravitational self-energy term does not represent the ill-definedness of time-translation operator in the strictly Newtonian regime (Christian 2001). In this regime, the time-translation operator can be well defined, but the gravitational self-energy term is obviously not zero. Moreover, as Di\'{o}si (2007) pointed out, the microscopic formulation of Penrose's collapse time formula also meets the cut-off difficulty.

Next, let's examine Penrose's choice of the preferred basis. According to Penrose (1998), the preferred bases are the stationary solutions of the Schr\"{o}dinger-Newton equation:

\begin{equation}
i\hbar {\partial \psi(\bold{x},t) \over \partial t}=-{\hbar^2 \over 2m}\nabla^2 \psi(\bold{x},t)-Gm^2\int{{|\psi(\bold{x'},t)|^2 \over |\bold{x}-\bold{x'}|}d^3 \bold{x'}}\psi(\bold{x},t)+V\psi(\bold{x},t),
\label{SN}
\end{equation}

\noindent  where $m$ is the mass of a quantum system, $V$ is an external potential, $G$  is Newton's gravitational constant\footnote{It has been shown that the Schr\"{o}dinger-Newton equation for spherically symmetric gravitational fields can be derived by WKB-like methods from the Einstein-Klein-Gordon and Einstein-Dirac system (Giulini and Gro$\beta$ardt 2012).}. This equation describes the gravitational self-interaction of a single quantum system, in which the mass density $m|\psi(\bold{x},t)|^2$ is the source of the classical gravitational potential. However, there is an obvious objection to the Schr\"{o}dinger-Newton equation (see also Giulini and Gro$\beta$ardt 2012). Since charge accompanies mass for a charged particle such as an electron, the existence of the gravitational self-interaction, though which is too weak to be excluded by present experiments (Salzman and Carlip 2006; Giulini and Gro$\beta$ardt 2011)\footnote{Note that Salzman and Carlip (2006) overestimated the influence of gravitational self-interaction on the dispersion of wave packets by about 6 orders of magnitude. This was pointed out and corrected by Giulini and Gro$\beta$ardt (2011).}, entails the existence of a remarkable electrostatic self-interaction of the charged particle, which seems incompatible with experiments. For example, for the electron in the hydrogen atom, the potential of the electrostatic self-interaction is of the same order as the Coulomb potential produced by the nucleus, and thus it seems impossible that the revised Schr\"{o}dinger equation with such an electrostatic self-interaction term, like the Schr\"{o}dinger equation, gives predictions of the hydrogen spectra that agree with experiment\footnote{However, since the revised Schr\"{o}dinger equation is essentially nonlinear, a strict analysis may be needed before a definite conclusion can be reached. For a more detailed discussion see Giulini and Gro$\beta$ardt (2012) and references therein.}.

On the other hand, it is worth noting that protective measurements show that a charged quantum system such as an electron does have mass and charge distributions in space, and the mass and charge density in each position is also proportional to the modulus squared of the wave function of the system there (Aharonov and Vaidman 1993; Aharonov, Anandan and Vaidman 1993; Gao 2011). However, the distributions do not necessarily exist throughout space at the same time, for which there are gravitational and electrostatic self-interactions of the distributions. Rather, they may be effective, which means that the distributions are formed by the ergodic motion of a point-like particle with the total mass and charge of the system. In this case, there will exist no gravitational and electrostatic self-interactions of the distributions, as there is only a localized particle at every instant. This is more consistent with the superposition principle and the Schr\"{o}dinger equation. It has been suggested that the wave function in quantum mechanics may represent the state of such motion of particles, which is arguably discontinuous and random in nature (Gao 2011).

Lastly, we briefly discuss three general problems of dynamical collapse models including Penrose's scheme\footnote{Pearle (2007, 2009), Bassi (2007) and Ghirardi (2011) presented a more detailed analysis of  the general problems of collapse models and the present status of the investigations of them.}. The first one is the origin of the randomness of collapse results. It is usually assumed, e.g. in the Continuous Spontaneous Localization (CSL) model (Pearle 1989; Ghirardi, Pearle and Rimini 1990), that the collapse of the wave function of a quantum system is caused by its interaction with an external noise field. Moreover, it has been suggested that the field is the background gravitational field, and the randomness of collapse results originates from the fluctuations of the gravitational field (see, e.g. K\'{a}rolyh\'{a}zy, Frenkel and Luk\'{a}cs 1986; Di\'{o}si 1989, 2007; Pearle and Squires 1996). However, it is worth noting that Penrose's gravity-induced collapse argument, even if it is valid, does not apply to these models, as the noise field in the models is not the gravitational field of the studied quantum system but the background gravitational field. It seems difficult to explain why the fluctuations of the background gravitational field have the extraordinary ability to cause the collapse of the wave function of a quantum system, though they may readily lead to the decoherence of the wave function of the system\footnote{In fact, since the Schr\"{o}dinger equation is purely deterministic, the quantum fluctuations must also result from the collapse of the wave function in these models. Thus it seems that these models are based on circular reasonings.}. On the other hand, if wavefunction collapse is spontaneous as in Penrose's scheme, then the randomness of collapse results cannot come from any external source, but must come from the studied quantum system itself. Yet the gravitational field of the studied quantum system seems to contain no such randomness\footnote{It has been suggested that the randomness may come from the random motion of particles described by the wave function (Gao 2011).}. 

Another problem of dynamical collapse models is energy non-conservation. For example, the collapse in the CSL model narrows the wave function in position space, therefore producing an increase of energy. A possible solution is that the conservation laws may be satisfied when the contributions of the external noise field to the conserved quantities are taken into account. It has been shown that the total mean energy can be conserved (Pearle 2000), and the energy increase can be made finite when revising the coupling between the noise field and the studied quantum system (Bassi, Ippoliti and Vacchini 2005). But it is still unclear whether energy can be strictly conserved in the model. As to Penrose's gravity-induced collapse scheme, although he did not give a concrete model of wavefunction collapse, he thought that the energy uncertainty $E_{\Delta}$ may cover such a potential non-conservation, leading to no actual violation of energy conservation (Penrose 2004). However, this is still a controversial issue. For instance, Di\'{o}si (2007) pointed out that the von-Neumann-Newton equation, which may be regarded as one realization of Penrose's scheme, does not conserve energy. On the other hand, there might also exist a possibility that the principle of conservation of energy is not universal and indeed violated by wavefunction collapse. One hint is that the usual proof that spacetime translation invariance leads to the conservation of energy and momentum relies on the linearity of quantum dynamics, and it does not apply to nonlinear quantum dynamics such as wavefunction collapse (Gao 2011). Moreover, such a violation of energy conservation may be so tiny that it is still consistent with present experiments.

The third problem is to make a relativistic quantum field theory which describes wavefunction collapse (Pearle 2009; Ghirardi 2011). In the CSL model, the hypothetical point interactions responsible for collapse will produce too many particles out of the vacuum and result in physically unacceptable divergent behaviour. It has been suggested that the problem of infinities may be solved by smearing out the point interactions. For example, Nicrosini and Rimini (2003) showed that this is possible when including a locally preferred frame. More recently, Bedingham (2011) introduced a new relativistic field responsible for mediating the collapse process, and showed that his model can fulfill the aim of smearing the interactions whilst preserving Lorentz covariance and frame independence. Whether this promising model is wholly satisfactory needs to be further studied. In addition, it is still unclear how to extend Penrose's scheme to the relativistic domain (cf. Anandan 1998).

In conclusion, we have argued that Penrose's proposal that gravity induces wavefunction collapse is debatable. However, it is still possible that wavefunction collapse is a real physical process as Penrose thinks, though its origin remains a deep mystery. Moreover, relating the process with gravity is still an extremely crucial problem which deserves a lot of attention, and approaches that are not fully satisfactory may also give hints concerning where to go or how to proceed.

\section*{Acknowledgments}
I am very grateful to two anonymous reviewers for insightful comments, constructive criticisms and helpful suggestions.

\section*{References}
\renewcommand{\theenumi}{\arabic{enumi}}
\renewcommand{\labelenumi}{[\theenumi]}
\begin{enumerate}

\item Anandan, J. S. (1998). Quantum measurement problem and the gravitational field, in S. A. Huggett, L. J. Mason, K. P. Tod, S. T. Tsou, and N. M. J.Woodhouse (eds.), The Geometric Universe: Science, Geometry, and
the Work of Roger Penrose. Oxford: Oxford University Press, pp. 357-368.
\item Anderson, E. (2012). The problem of time in quantum gravity,  In Classical and Quantum Gravity: Theory, Analysis and Applications, ed. V. R. Frignanni. New York: Nova Science. ch.4.
\item Bassi, A., Ippoliti, E., and  Vacchini, B. (2005). On the energy increase in space-collapse models. J. Phys. A : Math. Gen. 38, 8017.
\item  Bassi, A. (2007). Dynamical reduction models: present status and future developments. J. Phys.: Conf. Series 67, 012013.
\item  Bedingham, D. J. (2011). Relativistic state reduction dynamics. Found. Phys. 41, 686-704.
\item Christian, J. (2001). Why the quantum must yield to gravity. In: Physics Meets Philosophy at the Planck Scale, C. Callender and N. Huggett (ed.). Cambridge: Cambridge University Press. p. 305.
\item DeWitt, C. and Rickles, D. (ed.) (2011). The Role of Gravitation in Physics: Report from the 1957 Chapel Hill Conference. Max Planck Research Library for the History and Development of Knowledge, Vol. 5.
\item Di\'{o}si, L. (1984). Gravitation and the quantum-mechanical localization of macro-objects. Phys. Lett. A 105, 199-202.
\item Di\'{o}si, L. (1987). A universal master equation for the gravitational violation of quantum mechanics. Phys. Lett. A 120, 377-381.
\item Di\'{o}si, L. (1989). Models for universal reduction of macroscopic quantum fluctuations. Phys. Rev. A 40, 1165-1173.
\item Di\'{o}si, L.  (2007). Notes on certain Newton gravity mechanisms of wave function localisation and decoherence. J. Phys. A: Math. Gen. 40, 2989-2995.
\item Feynman, R. (1995). Feynman Lectures on Gravitation. B. Hatfield (ed.), Reading, Massachusetts:  Addison-Wesley.
\item Gao, S. (2010). On Di\'{o}si-Penrose criterion of gravity-induced quantum collapse. Int. J. Theor. Phys. 49, 849-853.
\item Gao, S. (2011). Interpreting quantum mechanics in terms of random discontinuous motion of particles. http://philsci-archive.pitt.edu/9057.
\item  Ghirardi, G. C. (2011). Collapse Theories. The Stanford Encyclopedia of Philosophy (Winter 2011 Edition), Edward N. Zalta (ed.), http://plato.sta-nford.edu/archives/win2011/entries/qm-collapse/.
\item  Ghirardi, G. C., Grassi, R., and Rimini, A. (1990). Continuous spontaneous reduction model involving gravity. Phys. Rev. A 42, 1057.
\item  Giulini, D. and Gro$\beta$ardt, A. (2011). Gravitationally induced inhibitions of dispersion according to the Schr\"{o}dinger-Newton equation. Class. Quant. Grav. 28, 195026.
\item  Giulini, D. and Gro$\beta$ardt, A. (2012). The Schr\"{o}dinger-Newton equation as non-relativistic limit of self-gravitating Klein-Gordon and Dirac fields. Class. Quant. Grav. 29, 215010.
\item Isham, C. J. (1993). Canonical Quantum Gravity and the Problem of Time, in L. A. Ibort and M. A. Rodriguez (eds.), Integrable Systems, Quantum Groups, and Quantum Field Theories. London: Kluwer Academic, pp. 157-288.
\item Isham, C. J. and Butterfield, J. (1999). On the emergence of time in quantum gravity, in The Arguments of Time, edited by J. Butterfield. Oxford: Oxford University Press. pp.111-168.
\item Kiefer, C. (2007). Quantum Gravity (Second Edition). Oxford: Oxford University Press. 
\item K\'{a}rolyh\'{a}zy, F. (1966). Gravitation and quantum mechanics of macroscopic objects. Nuovo Cimento A 42, 390-402.
\item K\'{a}rolyh\'{a}zy, F., Frenkel, A. and Luk\'{a}cs, B. (1986). On the possible role of gravity on the reduction of the wavefunction, in R. Penrose and C. J. Isham (eds.), Quantum Concepts in Space and Time. Oxford: Clarendon Press, pp. 109-128.
\item Kucha\v r, K. V. (1992). Time and interpretations of quantum gravity. In Proceedings of the 4th Canadian Conference on General Relativity and Relativistic Astrophysics (ed. G. Kunstatter, D. Vincent and J. Williams). Singapore: World Scientific.
\item Nicrosini, O. and Rimini, A. (2003). Relativistic Spontaneous Localization: A Proposal. Found. Phys. 33, 1061.
\item Pearle, P. (2000). Wavefunction Collapse and Conservation Laws. Found. Phys. 30, 1145-1160.
\item Pearle, P. (2007). How stands collapse I. J. Phys. A: Math. Theor., 40, 3189-3204.
\item Pearle, P. (2009). How stands collapse II. in Myrvold, W. C. and Christian, J.  eds., Quantum Reality, Relativistic Causality, and Closing the Epistemic Circle: Essays in Honour of Abner Shimony. The University of Western Ontario Series in Philosophy of Science, 73(IV), 257-292.
\item Pearle, P. and Squires, E. (1996). Gravity, energy conservation and parameter values in collapse models. Found. Phys. 26, 291.
\item Penrose, R.  (1981). Time-asymmetry and quantum gravity, in C. J. Isham, R. Penrose, and D. W. Sciama
(eds.), Quantum Gravity 2: a Second Oxford Symposium. Oxford: Oxford University Press, pp.
244-272.
\item Penrose, R. (1986). Gravity and state-vector reduction. in R. Penrose and C. J. Isham (eds.), Quantum Concepts in Space and Time. Oxford: Clarendon Press, pp. 129-146.
\item Penrose, R. (1989). The Emperor's New Mind: Concerning Computers, Minds, and the Laws of Physics. Oxford: Oxford University Press.
\item Penrose, R. (1994). Shadows of the Mind: An Approach to the Missing Science of Consciousness. Oxford: Oxford University Press.
\item Penrose, R. (1996). On gravity's role in quantum state reduction. Gen. Rel. Grav. 28, 581.
\item Penrose, R. (1998). Quantum computation, entanglement and state reduction. Phil. Trans. R. Soc. Lond. A 356, 1927.
\item Penrose, R. (2000). Wavefunction collapse as a real gravitational effect. In Mathematical Physics 2000 (ed. A. Fokas, T. W. B. Kibble, A. Grigouriou, and B. Zegarlinski), London: Imperial College Press. pp. 266-282. 
\item Penrose, R. (2002) Gravitational collapse of the wavefunction: an experimentally testable proposal. 
In Proceedings of the Ninth Marcel Grossmann Meeting on General Relativity (ed. V.G. Gurzadyan, R.T. Jantzen, R. Ruffini), Singapore: World Scientific. pp. 3-6.
\item Penrose, R. (2004). The Road to Reality: A Complete Guide to the Laws of the Universe. London: Jonathan Cape.
\item Rovelli, C.  (2004). Quantum Gravity, Cambridge: Cambridge University Press.
\item Rovelli, C.  (2011). ``Forget time": Essay written for the FQXi contest on the Nature of Time. Found. Phys. 41, 1475-1490.
\item Salzman, P. J. and Carlip, S. (2006). A possible experimental test of quantized gravity. arXiv: gr-qc/0606120.
\item Yablonovitch, E. (1987). Inhibited spontaneous emission in solid-state physics and electronics. Phys. Rev. Lett. 58, 2059.

\end{enumerate}
\end{document}